\documentclass[shortnote,twocolumn]{jpsj2}
\usepackage{graphicx}

\def\runtitle{a sample}
\def\runauthor{Jun-ichiro \textsc{Ohe}, Masami \textsc{Yamamoto},
Tomi \textsc{Ohtsuki} and Keith \texsc{Slevin}}

\title{%
Spin-dependent electron transport through a ferromagnetic domain wall
}

\author{%
Jun-ichiro \textsc{Ohe}
\thanks{E-mail:j-ohe@sophia.ac.jp}, Masami \textsc{Yamamoto},
Tomi \textsc{Ohtsuki} and Keith \textsc{Slevin}$^{1}$
}

\inst{%
Department of Physics, Sophia University, Japan\\
$^1$Department of Physics, Graduate School of Science, Osaka University, Japan
}

\recdate{\today}

\abst
{
}

\kword{%
domain wall, univesal conductance flactuation, spintronics
}

\begin{document}
\sloppy
\maketitle

Recent advances in spin-sensitive electronics made it possible
to realize spin-dependent electron transport in mesoscopic systems.
Utilizing the carrier's spin degrees of freedom in addition to their
charge opens the
possibilities of new spintronic devices \cite{ref1}.
The injection of spin polarized electrons in mesoscopic
systems can be achieved by a current from a ferromagnetic lead.
One of the most interesting transport properties in such systems
is the magnetoresistance induced by the electron scattering
by the spatial variation of the magnetization,
namely the domain wall, controlled by the external field.
The width of a domain wall can now be directly
measured and geometrically controlled \cite{ref2}.
For the application  to the actual devices,
it is important to extend the understanding of spin-dependent transport
through such a controllable domain wall system.
In this contribution,
we present a theoretical study of spin-dependent transport
through a ferromagnetic domain wall.
We calculate the conductance of this system by using the transfer matrix
method \cite{ref3} and extending it
to include the spin degrees of freedom.

We have studied the behavior of the conductance as a function of 
the geometry and strength of the domain wall.
We have considered three kinds of domain walls, namely
Ising, XY and Heisenberg types.
With an increase of the number of components of the exchange coupling,
we have observed that the variance of the conductance becomes half.
As the strength of the domain wall magnetization is increased,
negative magnetoresistance and a change of the conductance fluctuation
are observed. 

We consider a two dimensional (2D) system connected to two electrodes.
The 2D system is constructed in the $x$ and $y$ directions 
and current flows in the $x$ direction.
An exchange interaction between the electron spin and the static
local spin exists in the system.
The one-electron Hamiltonian is
\begin{eqnarray}
H=-\sum_{<i,j>}c^+_ic_j+\sum_iW_ic_i^+c_i-J\sum_ic_i^+\boldsymbol{\sigma}c_i
\cdot {\bf S}(x),
\end{eqnarray}
where $c^+_i (c_i)$ denotes the creation (annihilation) operator
of electron at the site $i$ on the 2D square lattice.
The transfer energy is taken to be the unit energy.
Energies $W_i$ denote the random potential distributed
independently and uniformly
in the range $[-W/2,W/2]$.
The hopping is restricted to nearest neighbors.
$\boldsymbol{\sigma}$ is the Pauli spin matrix and ${\bf S}(x)$ is the
local spin which has a spatial dependence of a domain wall.
Three types of domain walls, Ising, XY and Heisenberg,
are modelled as follows:
for the Ising domain wall we set
\begin{eqnarray}
S_x(x)=0, S_y(x)=0, \hspace{4.5cm}\nonumber \\
S_z(x)=\left\{
	\begin{array}{ll}
	S_0 \quad \mbox{for $x\le L$,} \nonumber \\
	S_0\cos (\frac{\pi(x-L)}{\lambda})\quad \mbox{for $L<x<L+N\lambda$,} \nonumber \\
	S_0\cos (N\pi ) \quad \mbox{for $x \ge L+N\lambda$,} \nonumber 	
	\end{array}
\right.
\end{eqnarray}
where $L$ is the position of the beginning of the domain wall,
$N$ the number of rotations and
$\lambda$ the length of the domain wall.
An XY domain wall is given by
\begin{eqnarray}
S_x(x)=0, \hspace{6.2cm}\nonumber \\
S_y(x)=\left\{
	\begin{array}{ll}
	S_0\sin (\frac{\pi(x-L)}{\lambda})
	\quad \mbox{for $L<x<L+N\lambda$,} \nonumber \\
	0 \quad \mbox{for otherwise,} \nonumber
	\end{array}
\right. \\
S_z(x)=\left\{
	\begin{array}{ll}
	S_0 \quad \mbox{for $x\le L$,} \nonumber \\
	S_0\cos (\frac{\pi(x-L)}{\lambda}) 
	\quad \mbox{for $L<x<L+N\lambda$,} \nonumber \\
	S_0\cos (N\pi ) \quad
	\mbox{for $x \ge L+N\lambda$.} \nonumber 	
	\end{array}
\right.
\end{eqnarray}
Similarly, a Heisenberg domain wall is modelled as
\begin{eqnarray}
S_x(x)=\left\{
	\begin{array}{ll}
	S_0\sin (\frac{\pi(x-L)}{\lambda})
	\quad \mbox{for $L<x<L+N\lambda$,} \nonumber \\
	0 \quad \mbox{for otherwise,} \nonumber
	\end{array}
\right. \hspace{2cm}\\
S_y(x)=\left\{
	\begin{array}{ll}
	S_0\cos (\frac{\pi(x-L)}{\lambda})\sin (\frac{\pi(x-L)}{\lambda}) 
	\quad \mbox{for $L<x<L+N\lambda$,} \nonumber \\
	0 \quad
	\mbox{for otherwise.} \nonumber 	
	\end{array}
\right.\hspace{0.2cm}\\
S_z(x)=\left\{
	\begin{array}{ll}
	S_0 \quad \mbox{for $x\le L$,} \nonumber \\
	S_0\cos^2 (\frac{\pi(x-L)}{\lambda}) 
	\quad \mbox{for $L<x<L+N\lambda$,} \nonumber \\
	S_0\cos (N\pi) \quad
	\mbox{for $x \ge L+N\lambda$.} &\nonumber 	
	\end{array}
\right.\hspace{1.5cm}
\end{eqnarray}
The conductance $G$ is given by Landauer's formula as
\begin{eqnarray}
G=\frac{e^2}{h} \mbox{tr}(tt^+) = \frac{e^2}{h} \sum_i \tau_i,
\end{eqnarray}
where $t$ is the transfer matrix\cite{ref3} including the spin degree of
freedom and $\tau_i$ the transmission eigenvalue.

\begin{figure}[t]
\begin{center}
\includegraphics[scale=0.85]{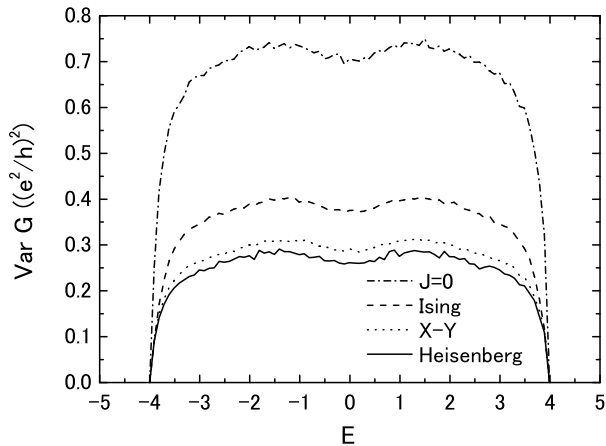}
\vspace{-1cm}
\caption{Variance of the conductance for $N=1$.
Other parameters are $W=3.0$, $\lambda=10$ and $L=1$.}
\end{center}
\end{figure}

\begin{figure}[t]
\begin{center}
\includegraphics[scale=0.85]{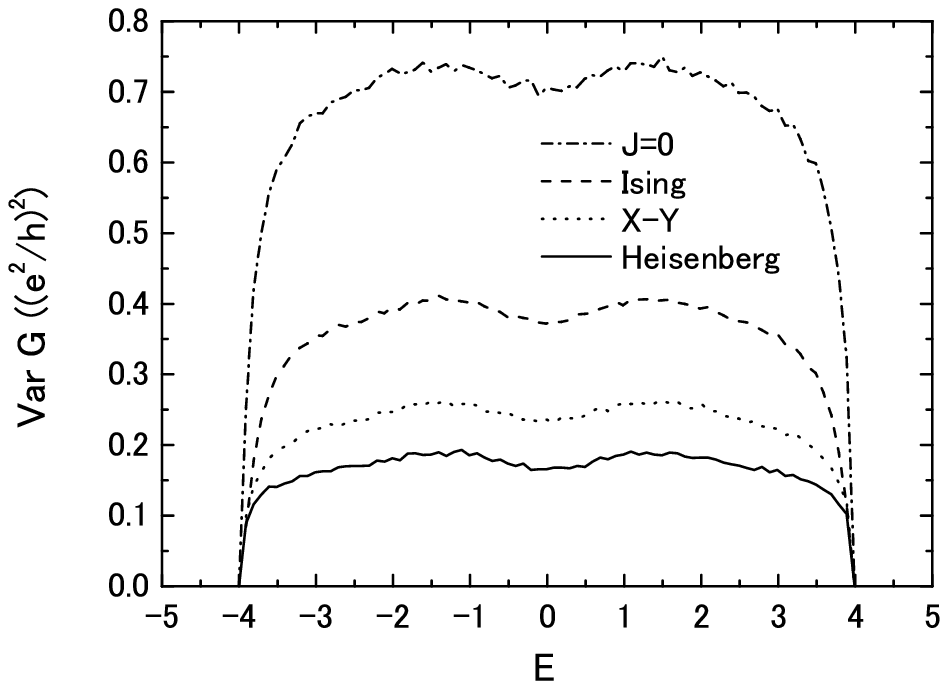}
\vspace{-1cm}
\caption{Variance of the conductance for $N=2$.
Other parameter is the same with Fig.~1.}
\end{center}
\end{figure}

\begin{figure}[t]
\begin{center}
\includegraphics[scale=0.85]{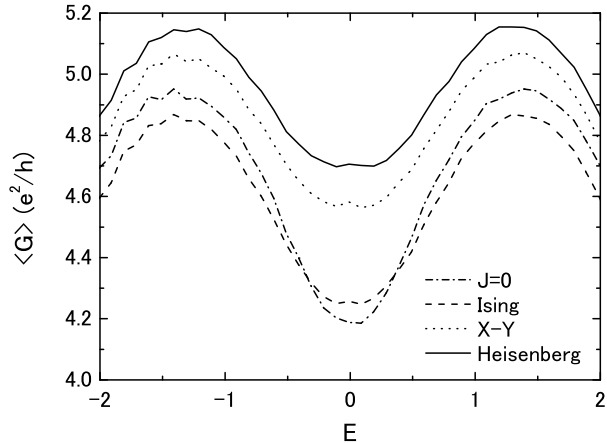}
\vspace{-1cm}
\caption{Average of the conductance for $N=2$.
Other parameter is the same with Fig.~1.}
\end{center}
\end{figure}

In the preset simulation, the system size is $30\times 30$
in units of lattice spacing, $W$ is set to be 3.0, and  at least
50,000 samples are taken for each ensemble.
$JS_0=\tilde{J}=3.0$ for the three types of domain walls, and
$\lambda=10$ and  $L=1$.

In Fig.~1 we show variance of conductance of these systems
with $N=1$.
When $\tilde{J}=0$,
there is no magnetic scattering and the variance
is close to that expected for
the universal conductance fluctuation of orthogonal systems.\cite{LSF}
For an Ising domain wall, spin-up and spin-down electrons are
described by different wave functions,
because the exchange field of domain wall breaks the spin degeneracy.
Therefore, the value of variance becomes half of the case for $\tilde{J}=0$.

While the Ising domain wall does not rotate the spin direction,
the spin flip process occurs in XY and Heisenberg
domain wall models.
In these cases, the variance of two systems are reduced by
the Ising domain wall case as shown in Fig.~1.

From the point of view of the symmetry of the Hamiltonian,
the XY domain wall model is classified into the orthogonal
class, while the Heisenberg domain wall model  is classified
into the unitary.
In Fig.~1, the difference of the variance of the conductance 
between XY and Heisenberg cases is small, but
for $N=2$ the reduction of the variance for the Heisenberg model
is prominent.

The results of the reduction of the variance are interpreted as follows.
The conductance fluctuation is determined by the spectral
statistics of the transmission eigenvalues $\tau$.
Let us denote the variance of $\sum_i{\tau_i}$ for orthogonal 2D
class as $V_{\rm 2DO}$ and the $G/(e^2/h)=\tilde{G}$.
Then due to the spin degeneracy in the case of $\tilde{J}=0$,
${\rm Var } ~\tilde{G}=4 V_{\rm 2DO}$.
For sufficiently strong Ising domain walls,
the up and down spin states form an independent ensemble and
${\rm Var } ~\tilde{G}=2 V_{\rm 2DO}$.
In the presence of XY domain walls, the variance is simply given by 
${\rm Var } ~\tilde{G}=V_{\rm 2DO}$.
Systems with Heisenberg domain walls belong to the
unitary class and ${\rm Var } ~\tilde{G}=\frac{1}{2} V_{\rm 2DO}$.

Our results also show that the domain walls
suppress the weak localization
effect due to impurity scattering.
Fig.~3 shows the average conductance.
System parameters are the same as in Fig.~2.
We observe  increases of conductance in the presence of
XY or Heisenberg types domain walls.



\section*{Acknowledgements}

We are grateful to T. Nakayama and K. Yakubo for valuable discussion.
One of the authors (J.O.) was supported by Research Fellowships of
the Japan Society for the Promotion of Science for Young Scientists.

%
%


\begin{thebibliography}{99}
\bibitem{ref1} S. Datta and B. Das; Appl. Phys. Lett. {\bf 56} (1990) 665. 
\bibitem{ref2} P. Bruno; Phys. Rev. Lett. {\bf 83} (1999) 2425. 
\bibitem{ref3} J. B. Pendry, A. MacKinnon and P. J. Roberts;
Proc. R. Soc. A {\bf 437} (1992) 67. 
\bibitem{LSF} P.A. Lee, A.D. Stone and H. Fukuyama;
Phys. Rev. B{\bf 35} (1987) 1039. 
\end{thebibliography}
\end{document}